\documentclass[aps,twocolumn,showpacs]{revtex4}
\usepackage{graphicx}
\usepackage{amsmath}
\usepackage{amsfonts}
\usepackage{color}
\newcommand{\tbox}[1]{\mbox{\tiny #1}}
\newcommand{\bra}{\left\langle}
\newcommand{\ket}{\right\rangle}
\providecommand{\abs}[1]{\lvert#1\rvert}

\begin{document}

\title{On the extremal spectral properties of random graphs} 

\author{C. T. Mart\'inez-Mart\'inez$^{1,2}$ and J. A. M\'endez-Berm\'udez$^3$}

\address{
$^1$Universidad Aut\'onoma de Guerrero, Centro Acapulco CP 39610, Acapulco de Ju\'arez, Guerrero, Mexico \\
$^2$Facultad de ciencias. Universidad Aut\'onoma Benito Ju\'arez de Oaxaca, Oaxaca de Ju\'arez CP 68120, Oaxaca, Mexico \\
$^3$Instituto de F\'{\i}sica, Benem\'erita Universidad Aut\'onoma de Puebla, Puebla 72570, Mexico
}

\date{\today} \widetext


\begin{abstract}
In this work, we study some statistical properties of the extreme eigenstates of the randomly-weighted 
adjacency matrices of random graphs. We focus on two random graph models: Erd\H{o}s-R\'{e}nyi (ER) 
graphs and random geometric graphs (RGGs).
Indeed, the adjacency matrices of both graph models are diluted versions of the Gaussian Orthogonal 
Ensemble (GOE) of random matrix theory (RMT), such that a transition from the Poisson Ensemble (PE)
and the GOE is observed by increasing the graph average degree $\langle k \rangle$.
First, we write down expressions for the spectral density in terms of  $\langle k \rangle$ for the regimes 
below and above the percolation threshold.
Then, we show that the distributions of both, the largest $\lambda_1$ and second-largest $\lambda_2$ 
eigenvalues approach the Tracy-Widom distribution of type 1 for $\langle k \rangle \gg 1$, while
$\langle \lambda_1 \rangle = \sqrt{2\langle k \rangle}$.
Additionally, we demonstrate that the distributions of the normalized distance between $\lambda_1$ 
and $\lambda_2$, the distribution of the ratio between higher consecutive eigenvalues 
spacings, as well as the distributions of the inverse participation ratios of the extreme 
eigenstates display a clear PE--to--GOE transition as a function of $\langle k \rangle$, so any
of these distributions can be effectively used to probe the delocalization transition of the graph 
models without the need of the full spectrum.
\end{abstract}

\maketitle

\section{Introduction}
The spectrum of a graph is the set of eigenvalues of the corresponding adjacency matrix. 
Within the study of spectral properties of graphs, a relevant direction is the analysis of extremal 
spectral quantities, such as the principal eigenvalue or largest eigenvalue 
$\lambda_1$~\cite{KS03,CLV03,FDBV01,CBP23,Allesina2012}.
The largest eigenvalue is a relevant quantity in various contexts, not only for graphs and networks,
as it is associated with important properties such as stability, dynamic response, and certain 
structural characteristics of the corresponding system. In mechanical and structural engineering, 
for example, knowing the largest eigenvalue of the stiffness or mass matrix (which is related to the 
highest natural frequency of the system) is important to avoid destructive 
resonances~\cite{BM78, P81, DZ09, F72}. In finance, the largest eigenvalue of the asset return 
covariance matrix informs about risk concentration and helps construct optimal portfolios~\cite{LCBP99, PGRAS99, BGLP12}. 
In physics, particularly in the study of disordered systems and quantum chaos, the behavior of the largest eigenvalue of a Hamiltonian is key to understanding phenomena such as Anderson 
localization and the distribution of energy levels in complex quantum systems~\cite{A58, BGS84, EM08, M04}.
In graphs and networks, the largest eigenvalue of the corresponding adjacency matrix provides a measure of the connectivity of the system. It can indicate structural characteristics such as hubs, communities, or the presence of dominant nodes \cite{GKK01, C97}. It also influences the determination of dynamic properties, such as the threshold for epidemic propagation or synchronization \cite{ADKMZ08,PC98,BP02}.
As for the Laplacian matrix, its largest eigenvalue carries complementary information. While the smallest non-zero Laplacian eigenvalue is commonly studied due to its connection to network robustness and communication efficiency, the largest Laplacian eigenvalue is also significant. It governs the upper bound of the spectrum and is linked to the network's ability to synchronize in dynamical processes. In particular, the ratio between the largest and the second-smallest Laplacian eigenvalues (the spectral gap) is a key factor in assessing synchronizability in systems of coupled oscillators~\cite{ADKMZ08}. Moreover, the largest Laplacian eigenvalue can reflect structural heterogeneity and can be used to bound quantities like the chromatic number or the isoperimetric number of the graph~\cite{C97}.
Thus, the spectrum of the adjacency matrix and the Laplacian provides relevant information about the structure and dynamics of complex networks, and the extreme eigenvalues are important in characterizing these properties.

The spectrum of the binary adjacency matrix of Erd\H{o}s-R\'{e}nyi graphs has been widely 
studied, see for example~\cite{FDBV01, SS18}. In the binary case, there is a significant difference 
between the statistical properties of the spectrum's bulk and the spectrum's edges.
It is known that the expected value of the largest eigenvalue $\lambda_1$ scales as 
$Np$ when $Np \gg \log(N)$, where $N$ is the number of nodes and $p$ the connection probability. 
This arises because the standard binary adjacency matrix of the Erd\H{o}s-R\'{e}nyi model resembles 
a random matrix 
whose average degree $\langle k \rangle = Np$ governs its spectral properties~\cite{KS03}. 

However, in this work, we consider randomly-weighted adjacency matrices, which makes previous results not directly applicable.
The random weights we incorporate into the graph's adjacency matrices follow a Gaussian distribution with mean
 zero and unit variance. Thus, these adjacency matrices can be interpreted as diluted versions of matrices 
 from the Gaussian Orthogonal Ensemble (GOE) of random matrix theory (RMT)~\cite{M04}. 
For the GOE, as $ N\to\infty$, the largest eigenvalue $\lambda_1$ follows the Tracy-Widom distribution of type 1, which is known to characterize the fluctuations of extremal eigenvalues of large random matrices~\cite{TW02}.

Thus, in this work, we focus on the study of spectral properties of graphs 
using tools from RMT, with particular emphasis on extremal spectral characteristics. 
Specifically, we explore some statistical properties of the largest eigenvalue $\lambda_1$.
For this purpose, we consider two models of random graphs: 
The Erd\H{o}s-R\'{e}nyi random graph model and the random geometric graph model.

This paper is organized as follows. In Section~\ref{ER}, we analyze the spectral properties of the Erd\H{o}s-R\'{e}nyi model, beginning with a brief description of the model, followed by a study of its spectral density, the behavior of the largest eigenvalue, and the characteristics of the corresponding eigenvectors. In Appendix A, we extend the analysis to the random geometric graph model, presenting analogous results. Finally, in Section~\ref{conclusions}, we summarize and discuss the main findings for both models.

\section{Erd\H{o}s-R\'{e}nyi model}
\label{ER} 

\subsection{Model}

An Erd\H{o}s-R\'{e}nyi (ER) graph, denoted by $G(N,p)$, is an undirected random graph with $N$ 
independent vertices connected with probability $p$. Given two vertices $u$ and $v$, $p$ is the 
probability of an edge from vertex $u$ to vertex $v$, so $p \in [0,1]$. When $p=0$, the graph 
consists of $N$ isolated vertices; when $p=1$, it becomes a complete graph. We can generate graphs 
between these two extremes by varying the value of $p$ between $0$ and $1$. It is important to 
note that a given pair of parameters $(N,p)$ represents an infinite set of random graphs. Therefore, 
calculating a given property for a single graph may not be representative of the entire model. Instead, 
we can obtain more 
relevant information by calculating the average of that property over an ensemble of random graphs 
characterized by the same pair of parameters $(N,p)$. Although this statistical approach is a common 
practice in RMT, it is not as common in graph theory; however, it has been recently applied to several 
random graph models~\cite{MMRS20, AHMS20, AMRS20, MMRS21, MAMRP15, PRRCM20, PM23}.

As already said above, in this work we study spectral properties of ER graphs. Moreover, to enable the comparison with standard results from RMT, we assign Gaussian random weights to the entries 
of the adjacency matrix $\mathbf{A}$ of $G(N,p)$ as
\begin{equation}
A_{ij}=\left\{
\begin{array}{cl}
\sqrt{2} \epsilon_{ii} \ & \mbox{for $i=j$}, \\
\epsilon_{ij} & \mbox{if there is an edge between vertices $i$ and $j$},\\
0 \ & \mbox{otherwise}.
\end{array}
\right.
\label{Aij}
\end{equation}
Here, we choose $\epsilon_{ij}$ as statistically independent random variables drawn from a normal 
distribution with zero mean and variance one. Also, $\epsilon_{ij}=\epsilon_{ji}$, since $G$ is 
undirected. According to this definition, diagonal random matrices are obtained for $p=0$ (Poisson 
ensemble (PE) in RMT terms), whereas the GOE (i.e.~full real and 
symmetric random matrices) is recovered when $p=1$~\cite{M04}. Therefore, a transition from the 
PE to the GOE can be observed by increasing $p$ from zero to one for any given graph size $N$.
Indeed, to characterize this PE--to--GOE transition, several RMT measures based on the eigenvalues 
and eigenvectors of the adjacency matrix of Eq.~(\ref{Aij}) have already been used; see e.g. 
Refs.~\cite{MAM15,TML18,TFM19,AMR20}.

In what follows, we call the ER random graphs represented by the randomly-weighted adjacency matrix 
of Eq.~(\ref{Aij}) as {\it randomly-weighted ER graphs.}

\subsection{Spectral Density}

We first examine the spectrum of randomly-weighted ER graphs by calculating the spectral density, 
which is the distribution of eigenvalues of the adjacency matrix. Indeed, for a finite graph, the spectral density can be expressed as a sum of Dirac delta functions: 
\begin{equation}
    \rho(\lambda):=\frac{1}{N}\sum_{j=1}^{N}\delta(\lambda-\lambda_{j}),
    \label{rho1}
\end{equation}
where $\lambda_j$ is the $j-$th eigenvalue of the graph's adjacency matrix.

A well-known result in RMT is Wigner's semicircle law~\cite{W58}. This law 
states that for a real, symmetric, $N \times N$ random matrix $A$ with uncorrelated elements, where 
$\langle A_{ij} \rangle =0$ and $\langle A_{ij}^{2} \rangle =\sigma^2$  for $i\neq j$, and where each 
moment of $\abs{A_{ij}}$ remains finite as $N$ increases, the spectral density of $A/\sqrt{N}$ 
converges to the semicircular distribution
\begin{equation}
\rho(\lambda)=\left\{
\begin{array}{cl}
(2 \pi \sigma^2)^{-1}\sqrt{4\sigma^2-\lambda^2} & \quad \mbox{if $\abs{\lambda}<2\sigma$},\\
0 \ & \quad \mbox{otherwise},
\end{array}
\right.
\label{rho2}
\end{equation}
in the limit $N \to \infty$.

Therefore, we expect the spectrum of randomly-weighted ER graphs to approach Wigner's semicircle 
law when $p\to 1$.
In contrast, notice that when $p\to 0$, the adjacency matrices are diagonal with random entries drawn 
from a Gaussian distribution, resulting in a spectral density following that Gaussian distribution.
Thus, in the transition from isolated to complete graphs, we expect to observe the transition from
a Gaussian spectral density to the Wigner's semicircle law.
For intermediate values of $p$, the average variance of the elements of the adjacency matrix is 
strongly affected by the number of elements equal to zero, which is closely related to the average degree of the graph.

The degree of a node is the number of edges connected to it. According to several studies,
see e.g.~\cite{MAM15,AHMS20, AMRS20, MMRS20,MMRS21, PRRCM20, PM23,MAMRP15}, the 
average degree is a quantity that characterizes several spectral and topological properties
of graphs and networks. In the case of ER-type graphs, the average degree is given by
\begin{equation}
    \langle k \rangle = (N-1)p.
\end{equation}
However, in our specific case, the graphs include self-loops, which means that even when the connection probability $p$ is zero, each node is still connected to itself. This results in a modified expression for the average degree:
\begin{equation}
    \langle k \rangle =1+ (N-1)p.
\end{equation}

The random weights of the adjacency matrix $\mathbf{A}$ have unit variance. 
However, for decreasing connection probability, the number of zero entries in the adjacency 
matrix increases, effectively reducing the overall variance of the matrix. This variance reduction 
can be directly connected to the average degree $\langle k \rangle$.
In particular, since the model includes self-connections, the percolation threshold is reached at 
$\langle k \rangle = 2$; at this point, connected components emerge more frequently. 
Specifically, we find that $\sigma = \langle k \rangle$ for $\langle k \rangle < 2$, while 
$\sigma = \langle k \rangle / 2$ for $\langle k \rangle > 2$.

So, following Wigner's law and considering the relationship between variance and average degree, we propose the following expression for the spectral density of randomly-weighted ER graphs:
\begin{equation}
\rho(\lambda)=\frac{1}{\sqrt{2\pi\langle k \rangle}} \exp\left({-\frac{\lambda^2}{2\langle k \rangle}}\right) ,
\label{den1}
\end{equation}
if $\langle k \rangle<2$, while 
\begin{equation}
\rho(\lambda)=\left\{
\begin{array}{cl}
(\pi \langle k \rangle )^{-1}\sqrt{2\langle k \rangle-\lambda^2} & \quad \mbox{if $\abs{\lambda}<\sqrt{2\langle k \rangle}$},\\
0 \ & \quad \mbox{otherwise},
\end{array}
\right.
\label{den2}
\end{equation}
if $\langle k \rangle>2$.
In Fig.~\ref{Fig01ER}, we present the spectral density of randomly-weighted ER graphs for several 
values of $\langle k \rangle$, as specified in the caption. Indeed, Eqs.~(\ref{den1}) (blue lines) 
and~(\ref{den2}) (red lines) fit the distributions well for small and large values of $\langle k \rangle$, 
respectively.

\renewcommand{\thefigure}{ERGs\arabic{figure}}
\setcounter{figure}{0}

\begin{figure}
\centering
\includegraphics[scale=0.4]{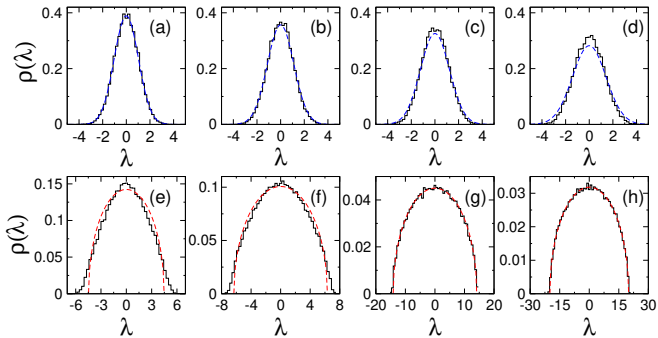}
\caption{
Spectral density of randomly-weighted Erd\H{o}s-R\'{e}nyi graphs of size $N=200$ 
and (a) $\langle k \rangle =1.05$, (b) $\langle k \rangle =1.25$, (c) $\langle k \rangle =1.5$,
(d) $\langle k \rangle =2$, (e) $\langle k \rangle =10$, (f) $\langle k \rangle =20$, 
(g) $\langle k \rangle =100$, and (h) $\langle k \rangle =200$. 
Blue and red dashed lines correspond to Eqs.~(\ref{den1}) and~(\ref{den2}), respectively. 
The eigenvalues of 100 adjacency matrices were used to construct each histogram.}
\label{Fig01ER}
\end{figure}

From Fig.~\ref{Fig01ER} we observe that, unlike the binary case, for randomly-weighted ER graphs
there is no clear separation between the largest and second-largest eigenvalues. Furthermore, for 
large values of $\langle k \rangle$, the largest eigenvalue appears to be directly related to the radius 
of the spectral density. Next, we analyze the behavior of the largest eigenvalue as a function of the 
model parameters.

\subsection{Largest eigenvalue}

In Figs.~\ref{Fig02ER}(a) and~\ref{Fig02ER}(b) we plot the average largest eigenvalue $\bra \lambda_{1}\ket$
of randomly-weighted ER graphs as a function of $p$ and $N$, respectively.
In the inset of Fig.~\ref{Fig02ER}(a), we present $\bra \lambda_{1}\ket$ as a function of the average degree
and observe that it depends solely on it. Moreover, according to Eq.~(\ref{den2}), the largest eigenvalue can 
be expressed as 
\begin{equation}
    \lambda_{1}=\sqrt{2\langle k \rangle}.
    \label{lmb1}
\end{equation}
The inset of Fig.~\ref{Fig02ER}(a) shows that Eq.~(\ref{lmb1}) (dashed black line) provides a good description 
of the numerical data for $k > 2$, which works even better for $k > 10$ (brown dashed line).
Fig.~\ref{Fig02ER}(b) shows the average largest eigenvalue $\langle \lambda_1 \rangle$ as a function of the network size $N$ for different values of the connection probability $p$, as indicated in the legend. Green symbols correspond to higher values of $p$, while red symbols indicate lower values. We can see that $\langle \lambda_1 \rangle$ follows a power-law dependence on $N$.
Based on Eq. (\ref{lmb1}), by substituting the expression for the average degree, we obtain the approximation $\lambda_1 \approx \sqrt{2Np}$. This approximation is indicated by dashed lines in the figure, particularly for higher values of $p$. We observe that this approximation agrees better with the numerical results when $p$ is bigger.

\begin{figure}
\centering
\includegraphics[scale=0.4]{Fig02ER.eps}
\caption{
(a) The average largest eigenvalue $\bra \lambda_{1}\ket$, of randomly-weighted Erd\H{o}s-R\'{e}nyi graphs,
as a function of the connection probability $p$. 
Inset: $\bra \lambda_{1}\ket$ as a function of the average degree $\langle k \rangle$.  
(b) $\bra \lambda_{1}\ket$ as a function of the graph size 
$N$ for different values of $p$, as indicated in the panel. 
Dashed lines in the inset of panel (a) and in panel (b) correspond to Eq.~(\ref{lmb1}).
Each symbol was calculated by averaging over $10^6/N$ random graphs.}
\label{Fig02ER}
\end{figure}

Moreover, we also explore the distribution of $\lambda_1$ for different parameter combinations. 
To ease the comparison of results, we normalize the largest eigenvalue as:
\begin{equation}
  \tilde{\lambda}_{1}=\frac{\lambda_1 -\langle \lambda_{1} \rangle }{\sigma_{\lambda_{1}}}  
  \label{nrei}
\end{equation}
where $\langle \lambda_1 \rangle$ is the mean value of the largest eigenvalue and $\sigma_{\lambda_1}$ its standard deviation.

\begin{figure}
\centering
\includegraphics[scale=0.4]{Fig03ER.eps}
\caption{
(a) The average variance of the largest eigenvalue $\bra \sigma_{1}\ket$, of randomly-weighted 
Erd\H{o}s-R\'{e}nyi graphs, as a function of the connection probability $p$. (b) $\bra \sigma_{1}\ket$ 
as a function of the graph size $N$ for different values of $p$, as indicated in the panel. 
Dashed lines in (a) correspond to $N^{-1/6}$ for $N=100,400$ and $1600$. 
Dashed line in (b) is $N^{-1/6}$
Each symbol was calculated by averaging over $10^6/n$ random graphs.}
\label{Fig03ER}
\end{figure}

Figure~\ref{Fig03ER} shows the average variance of the largest eigenvalue $\bra \sigma_{\lambda_{1}}\ket$ of 
randomly-weighted ER graphs as a function of (a) $p$ and (b) $N$.  From Fig.~\ref{Fig03ER}(a) we can see that $\bra \sigma_{\lambda_{1}}\ket$ 
is not a simple function of $p$: For small $p$ it remains almost constant, then decreases as $p$ 
grows, reaching a minimum, and finally increases with $p$.
As a function of the graph size $N$, see Fig.~\ref{Fig03ER}(b)), $\bra \sigma_{\lambda_{1}}\ket$ follows a 
decreasing power law, where the power law depends on the value of $p$.

For the GOE, when $N\to\infty$, the largest normalized eigenvalue $\tilde{\lambda}_1$ follows a 
distribution that converges to the Tracy-Widom distribution of type 1~\cite{TW02}.
In fact, for the GOE, the asymptotic expected value of the largest eigenvalue is given by 
$\langle \lambda_{1} \rangle = \sqrt{2N}$ and the scale of fluctuations around this expected value 
is given by $\sigma_{\lambda_{1}}=N^{-1/6}$ for large matrices;  this value has been plotted in 
dotted lines in Fig.~\ref{Fig03ER}(a) for $n=100$, $400$ and $1600$. 
More specifically, if $ \lambda_1$ is the largest eigenvalue of a GOE $N \times N$ matrix, then the
distribution of
\begin{equation}
  \tilde{\lambda}_{1}=\frac{\lambda_1 -\sqrt{2N} }{N^{-1/6}}  
\end{equation}
converges to the Tracy-Widom distribution of type 1. 

\begin{figure}
\centering
\includegraphics[scale=0.4]{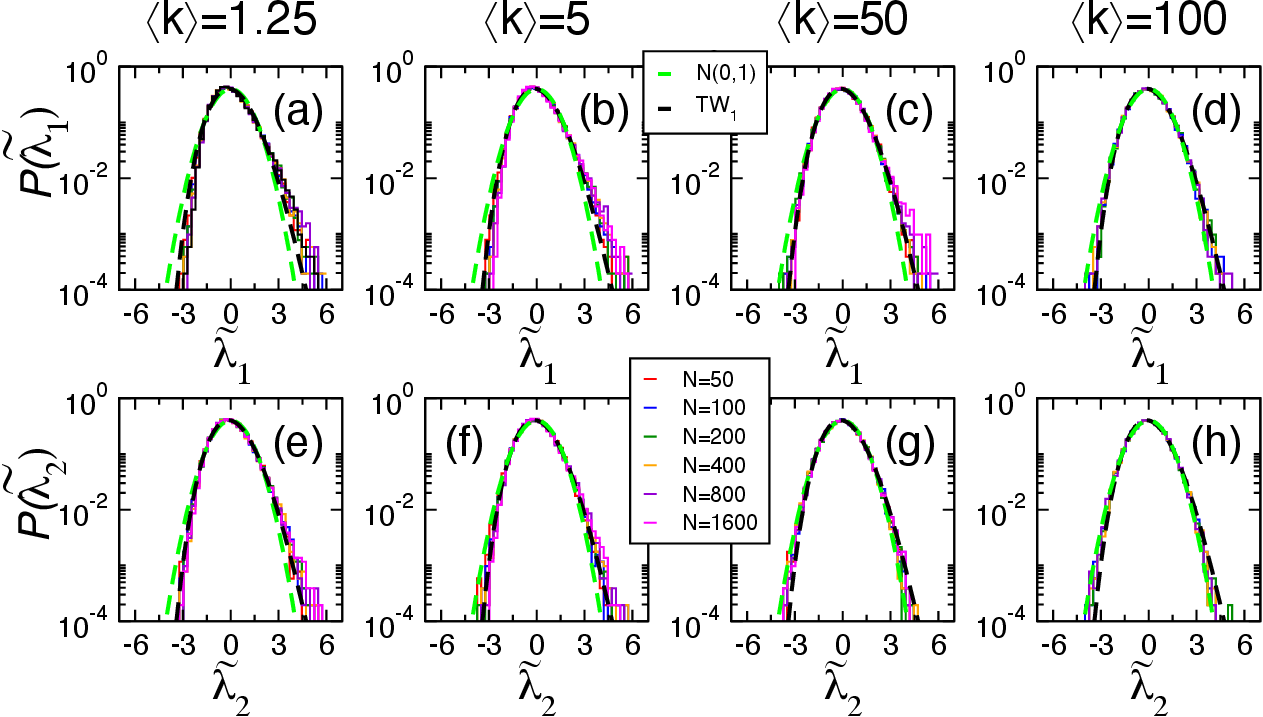}
\caption{
Top panels: Distributions of the largest eigenvalue $P(\tilde{\lambda}_1)$ of randomly-weighted 
Erd\H{o}s-R\'{e}nyi graphs for different values of $\langle k \rangle$, as indicated at the 
top of the panels. 
Bottom panels: Distributions of the second largest eigenvalue $P(\tilde{\lambda}_2)$.
Each panel includes histograms for several graph sizes $n \in [50,400]$. 
Black and green dashed lines correspond to the Tracy-Widom distribution of 
type 1 and normal distributions, respectively. 
Each histogram was constructed from $10^6/n$ random graphs.}
\label{Fig04ER}
\end{figure}

Then, in Fig.~\ref{Fig04ER} (top panels), we present distributions of the largest eigenvalue 
$P(\tilde{\lambda}_1)$ of randomly-weighted ER graphs. $\lambda_1$ is normalized according with Eq.~(\ref{nrei}).
Given that the average degree serves as a scaling parameter of several structural and spectral 
properties of graphs~\cite{AMRS20, PM23, MAM15, MMS24}, we decided to examine $P(\tilde{\lambda}_1)$ for fixed 
values of $\langle k \rangle$. To this end, we choose values of $\langle k \rangle$ from weakly to highly 
connected graphs: $\langle k \rangle =1.25$, 5, 50, and 100; as indicated on top of the panels.
In addition, in Fig.~\ref{Fig04ER} (lower panels), we present distributions of the second-largest eigenvalue. 
We note that the distributions of both (normalized) eigenvalues tend to the Tracy-Widom distribution of type 
1 for increasing $\langle k \rangle$, see the green dashed lines in Fig.~\ref{Fig04ER}(d,h).

\begin{figure}
\centering
\includegraphics[scale=0.3]{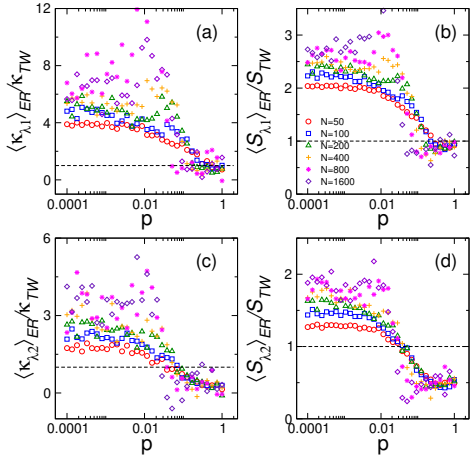}
\caption{
(a) [(c)] Excess kurtosis and (b) [(d)] skewness, normalized to the corresponding values of the 
Tracy-Widom distribution of type 1, of the distribution of the largest eigenvalue $\lambda_1$ 
[of the distribution of the 
second largest eigenvalue $\lambda_2$] as a function of the connection probability $p$ of 
randomly-weighted Erd\H{o}s-R\'{e}nyi graphs of sizes $n \in [50,400]$. 
Each symbol was computed by averaging over $10^6/n$ random graphs.}
\label{Fig05ER}
\end{figure}

To better characterize the overall behavior of the eigenvalue distributions of Fig.~\ref{Fig04ER}, 
we compute their excess kurtosis and skewness. Then, in Figs.~\ref{Fig05ER}(a) and~\ref{Fig05ER}(b), 
respectively, we plot the excess kurtosis and the skewness of the distribution of the largest eigenvalue 
$\lambda_1$ as a function of the connection probability $p$ of randomly-weighted ER graphs.
Both quantities are normalized to the corresponding values of the Tracy-Widom distribution of type 1,
which is approached when $p\to 1$; see Fig.~\ref{Fig04ER}.
In Figs.~\ref{Fig05ER}(c) and~\ref{Fig05ER}(d) we also plot the excess kurtosis and the skewness of
$P(\lambda_2)$, respectively.

From Figs.~\ref{Fig05ER}(a) and~\ref{Fig05ER}(b), when $p$ is small, we observe that the excess 
kurtosis and skewness of $P(\lambda_1)$ deviate considerably from the Tracy-Widom value. However, 
as $p$ increases, as expected, both quantities progressively approach the reference Tracy-Widom 
values; although fluctuations persist at large values of $p$.
In the case of $P(\lambda_2)$, the situation is different: While both the excess kurtosis and the skewness
decrease with $p$, they do not approach the reference Tracy-Widom values; see Figs.~\ref{Fig05ER}(c) 
and~\ref{Fig05ER}(d). This may be expected, even with the apparent good correspondence between 
$P(\lambda_2)$ and the Tracy-Widom distribution shown in Fig.~\ref{Fig04ER}(h), since Tracy-Widom--type
statistics is expected for $\lambda_1$ only.

\begin{figure}
\centering
\includegraphics[scale=0.3]{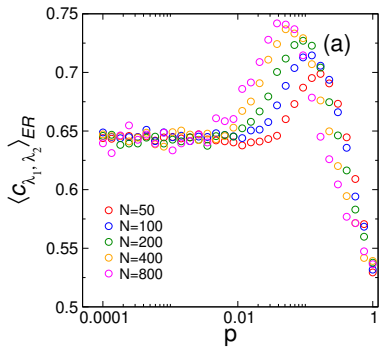}
\caption{
Correlation coefficient between the largest and the second largest eigenvalues 
as a function of the connection probability $p$ of randomly-weighted 
Erd\H{o}s-R\'{e}nyi graphs of sizes $n \in [50,400]$.
Each symbol was computed by averaging over $10^6/n$ random graphs.}

\label{Fig06ER}
\end{figure}

Now, we also compute the correlation coefficient between the first and second eigenvalues, 
$c_{\lambda_1,\lambda_2}$, defined by
\begin{equation}
    c_{\lambda_1,\lambda_2}=\langle \tilde{\lambda}_1\tilde{\lambda}_2\rangle-\langle\tilde{\lambda}_1\rangle\langle\tilde{\lambda}_2\rangle.
\end{equation}
$c_{\lambda_1,\lambda_2}$ as a function of $p$ of randomly-weighted ER graphs is shown
in Fig.~\ref{Fig06ER}. From this figure, we observe that, for small $p$, the correlation coefficient
remains approximately constant and close to $c_{\lambda_1,\lambda_2} = 0.65$. This may be
regarded as the PE value. Then, for further increasing $p$, $c_{\lambda_1,\lambda_2}$ grows,
reaches a maximum, and finally decreases until reaching the predicted value for the GOE, $c_{\lambda_1,\lambda_2} \approx 0.53$~\cite{CBP23}.

From the results presented so far, we have observed that the statistical properties of the largest eigenvalue
(and also of the second largest eigenvalue) of the model of randomly-weighted ER graphs shows a clear 
PE--to--GOE transition as a function of the graph connectivity, or more precisely, as a function of the average 
degree. 
Although the PE--to--GOE transition as a function of $\langle k \rangle$ has been reported previously by the 
use of the full spectrum, see 
e.g.~\cite{MAMRP15}, it is important to highlight that our results demonstrate that the PE--to--GOE transition 
could also be probed by the use of the extreme eigenvalues only. 
To further verify this statement, we now compute the distribution of the normalized distance between the largest 
and second-largest eigenvalues, defined as
\begin{equation}
    s=\frac{\lambda_{1}-\lambda_{2}}{\langle \lambda_{1}-\lambda_{2}\rangle},
\end{equation}
and the distribution of the ratio between higher consecutive eigenvalues spacings
\begin{equation}
    r=\frac{\min(\lambda_{1}-\lambda_{2},\lambda_{2}-\lambda_{3})}{\max(\lambda_{1}-\lambda_{2},\lambda_{2}-\lambda_{3})}.
\end{equation}

In Fig.~\ref{Fig07ER} we present $P(s)$ (top panels) and $P(r)$ (bottom panels) of randomly-weighted ER 
graphs for different values of $\langle k \rangle$, as indicated at the top of the panels.
Note that we are using the same values of $\langle k \rangle$ reported in Fig.~\ref{Fig04ER}.
In Fig.~\ref{Fig07ER} we are also including the RMT predictions of both $P(s)$ and $P(r)$ for the PE
and the GOE, which are given by
\begin{equation}
P_{\tbox{PE}}(s)=\exp{(-s)},
\label{psp}
\end{equation}
\begin{equation}
P_{\tbox{PE}}(r)=\frac{2}{(1+r)^2},
\label{prp}
\end{equation}
 \begin{equation}
P_{\tbox{GOE}}(s)=\frac{\pi}{2}s\exp{\left(-\frac{\pi}{4}s^2\right)},
\label{psg}
\end{equation}
and 
\begin{equation}
P_{\tbox{GOE}}(r)=\frac{27}{4}\frac{(r+r^2)}{(1+r+r^2)^{5/2}}.
\label{prg}
\end{equation}
However, it is important to stress that Eqs.~(\ref{psp}-\ref{prg}) are expected to work at the bulk of the
spectrum.
Then, from Fig.~\ref{Fig07ER} we observe that:
(i) for fixed $\langle k \rangle$ the shape of both $P(s)$ and $P(r)$ remains invariant, i.e.~they do not depend 
on the graph size; except for intermediate values of $\langle k \rangle$, see Figs.~\ref{Fig07ER}(b,f);
(ii) for small [large] values of $\langle k \rangle$, the shapes of $P(s)$ and $P(r)$ are well described by 
$P_{\tbox{PE}}(s)$ and $P_{\tbox{PE}}(r)$ [$P_{\tbox{GOE}}(s)$ and $P_{\tbox{GOE}}(r)$], respectively;
(iii) both $P(s)$ and $P(r)$ can be used to probe the PE--to--GOE transition.

\begin{figure}
\centering
\includegraphics[scale=0.4]{Fig07ER.eps}
\caption{
Top panels: Distribution of the normalized distance between the largest 
and second-largest eigenvalues $P(s)$ of randomly-weighted Erd\H{o}s-R\'{e}nyi graphs 
for different values of $\langle k \rangle$, as indicated at the top of each panel.
Bottom panels: Distribution of the ratio between higher consecutive eigenvalues spacings $P(r)$. 
Each panel displays histograms for different graph sizes, $n \in [50,400]$. 
In the top panels, black and green dashed lines correspond to Eqs.~(\ref{psp}) and~(\ref{psg}), respectively. 
In the bottom panels, black and green dashed lines correspond to Eqs.~(\ref{prp}) and~(\ref{prg}), respectively. 
Each histograms was constructed from $10^6/n$ random graphs.}
\label{Fig07ER}
\end{figure}

\subsection{Eigenvector properties}

\begin{figure}
\centering
\includegraphics[scale=0.4]{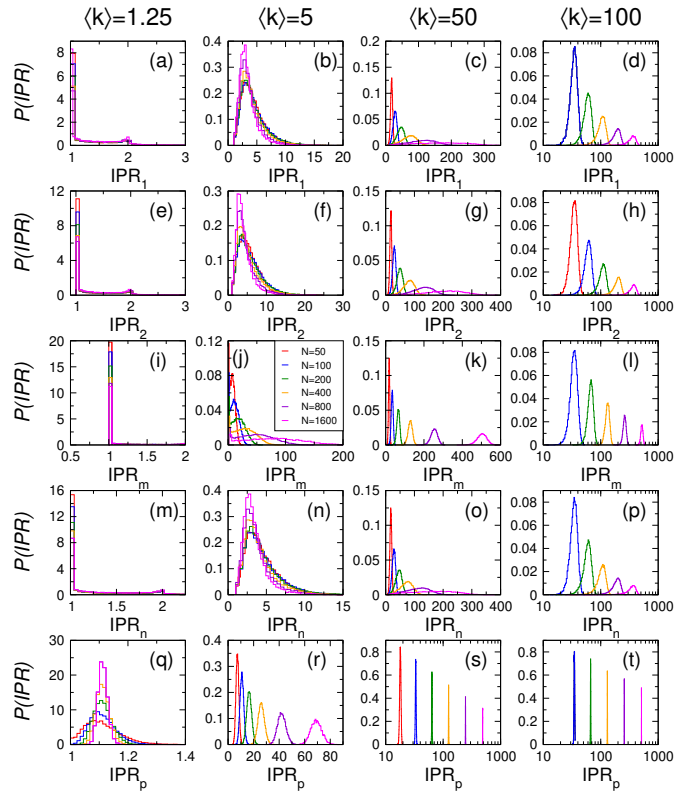}
\caption{
Distributions of inverse participation ratios $P(\mbox{IPR})$ of eigenstates of randomly-weighted 
Erd\H{o}s-R\'{e}nyi graphs of sizes $N \in [50,800]$. 
Each column corresponds to a fixed value of $\langle k \rangle$.
Rows are $P(\mbox{IPR})$ of eigenstates corresponding to: 
(a-d) the largest eigenvalue, (e-h) the second largest eigenvalue, 
(i-l) the central eigenvalue $(m=N/2)$, (m-p) the smallest eigenvalue, and (q-t) the average IPR 
over the full spectrum.
}
\label{Fig08ER}
\end{figure}

To further explore the properties of the edge of the spectrum, here we compute the inverse participation 
ratio (IPR) of the eigenvectors corresponding to the extreme eigenvalues.
Given the normalized eigenvectors \( \Psi^i \), its IPR is defined as
\begin{equation}
    \mbox{IPR}_{i}={\left[ \sum_{m=1}^{N} |\Psi_{m}^{i}|^{4}\right]}^{-1}.
\end{equation}
In fact, we compute distributions of IPRs of eigenvectors of randomly-weighted ER graphs, as shown in 
Fig.~\ref{Fig08ER}.
Notice that each column in Fig.~\ref{Fig08ER} corresponds to a fixed value of $\langle k \rangle$;
we used the same values of $\langle k \rangle$ reported in Figs.~\ref{Fig04ER} and~\ref{Fig07ER}.
Specifically, in Fig.~\ref{Fig08ER} we show the distributions of the IPRs of the eigenvectors corresponding 
to the largest eigenvalue, see panels (a-d), and to the second largest eigenvalue, see panels (e-h).
Moreover, for comparison purposes, we also show $P(\mbox{IPR})$ of the eigenvectors 
corresponding to the central eigenvalue, see panels (i-l), the smallest eigenvalue, see panels (m-p), 
and the average IPR over the full spectrum, see panels (q-t).

From Fig.~\ref{Fig08ER} we observe that all IPR distributions display a clear delocalization transition 
as the average degree of the graph increases.
For small values of $\langle k \rangle$, see the left panels corresponding to $\langle k \rangle = 5$, 
the IPR distributions show a pronounced peak at $\mbox{IPR}=3$, a signature of strongly localization and 
a characteristic of the PE.
For large values of $\langle k \rangle$, see the right panels corresponding to $\langle k \rangle = 100$, 
the IPR distributions follow a Gaussian-like bell shape centered at $\mbox{IPR}\propto N$, indicating 
delocalization which is the main characteristic of the GOE.
For intermediate values of $\langle k \rangle $, we observe a transition between the PE and the GOE regimes.
The eigenvectors corresponding to the extreme eigenvalues (largest, second largest, and smallest) exhibit 
similar localization characteristics. This, in contrast with the eigenvector corresponding to the central 
eigenvalue $\lambda_{N/2}$, which also displays the delocalization transition but in a different manner.
The distributions of the average IPR, shown in panels (q-t), included for comparison purposes, clearly show that, on average, the localization properties of the bulk eigenvalues dominate.

We recall then that the localization properties of the eigenvectors are not uniform across the spectrum. 
In general, the eigenvectors at the center of the spectrum are more delocalized than those at the edges.
Also, extreme states tend to be more closely aligned with the topological structure of the graph than those 
at the spectrum center, see e.g.~\cite{FDBV01}.
Indeed, the spectral heterogeneity is key to understanding dynamical processes in graphs, such as diffusion, 
synchronization, or stability, where different modes can have differentiated contributions depending on their 
degree of localization; see e.g.~\cite{ADKMZ08, PC98}.

\section{Conclusions and discussion}
\label{conclusions}

In this paper, we have studied some statistical properties of the extreme eigenstates of  
randomly-weighted adjacency matrices $\mathbf{A}$ corresponding to two random graph models: 
Erd\H{o}s-R\'{e}nyi (ER) graphs and random geometric graphs (RGGs).
Since we added self-connections to the graph's nodes, see Eq.~(\ref{Aij}), the adjacency matrices 
of both graph models 
are diluted versions of the Gaussian Orthogonal Ensemble (GOE) of random matrix theory (RMT), 
such that a transition from the Poisson Ensemble (PE) and the GOE is observed by increasing the 
graph average degree $\langle k \rangle$.
We note that, to avoid saturation of the main text, the results for randomly-weighted RGGs are
reported in the Appendix.

First, we wrote down expressions for the spectral density in terms of  $\langle k \rangle$ for the regimes 
below and above the percolation threshold; see Eqs.~(\ref{den1}) and~(\ref{den2}), respectively, and
Figs.~\ref{Fig01ER} and~\ref{Fig09RG}.
Then, we showed that the average value of the largest eigenvalue $\langle \lambda_1 \rangle$
depends on $\langle k \rangle$ only and it is well described by Eq.~(\ref{lmb1}); see the insets in
Figs.~\ref{Fig02ER}(a) and~\ref{Fig10RG}(a). Moreover, we verified that both distributions,
$P(\lambda_1)$ and $P(\lambda_2)$ (the distribution of the second-largest eigenvalue), approach 
the Tracy-Widom distribution of type 1 for $\langle k \rangle \gg 1$; see 
Figs.~\ref{Fig04ER} and~\ref{Fig12RG}.

We also demonstrated that the distributions of the normalized distance between $\lambda_1$ 
and $\lambda_2$, $P(s)$, the distribution of the ratio between higher consecutive eigenvalues 
spacings, $P(r)$, as well as the distributions of the inverse participation ratios of the extreme 
eigenstates display a clear PE--to--GOE transition as a function of $\langle k \rangle$, so any
of these distributions can be effectively used to probe the delocalization transition of the graph 
models without the need of the full spectrum. Moreover, for small [large] values of $\langle k \rangle$, 
the shapes of $P(s)$ and $P(r)$ are well described by the RMT predictions $P_{\tbox{PE}}(s)$ 
and $P_{\tbox{PE}}(r)$ [$P_{\tbox{GOE}}(s)$ and 
$P_{\tbox{GOE}}(r)$], respectively; see Figs.~\ref{Fig07ER} and~\ref{Fig15RG}.

These findings demonstrate the relevance of extreme spectral statistics as indicators of transitions in complex networks. In particular, they suggest that important information about the connectivity regime and overall system behavior can be obtained from just a few spectral observables. This is particularly beneficial in problems where accessing the full spectrum is computationally expensive. Future research could explore whether these indicators remain effective in characterizing other types of networks with more specific structures, as well as their relevance to real-world networks and dynamics, such as diffusion, synchronization, or signal propagation.

\appendix
\renewcommand{\thefigure}{RGGs\arabic{figure}}
\setcounter{figure}{0}

\section{Random geometric graphs}

In this Appendix, we report the statistical properties of the extreme states of the randomly-weighted 
adjacency matrices of random geometric graphs (RGGs).

The model of RGGs is defined as follows: $N$ nodes are distributed uniformly over a square with 
unit side. Then, if the distance between two nodes is less than a connection radius $r$, an edge is 
set between them. When $r=0$ the graph is completely disconnected, while if $r=\sqrt{2}$ the graph
is complete. This model is then defined by the parameters $N$, the total number of nodes or size of 
the graph, and $r$, the connection radius.

For RGGs, the expression for average degree is more complex than that for ER graphs. For RGGs, 
without self-connections, the average degree is given by~\cite{ES15}:
\begin{equation}
    \langle k \rangle =(N-1)F(r) ,
\end{equation}
where 
\begin{equation}
F(r)=\left\{ 
\begin{array}{ll}
\pi r^2-\frac{4}{3}r^3+\frac{1}{2}r^4  \ & \mbox{for $0<r\leq 1$}, \\ \ \\
\frac{1}{3}-\frac{1}{2}r^4+(\frac{8}{3}r^2+\frac{4}{3})\sqrt{r^2-1} \\  
-2r^2\left( \arccos{(\frac{1}{r})}-\arcsin{(\frac{1}{r})}+1\right)
  & \mbox{for $r>1$}.
\end{array}
\right.
\label{krgg}
\end{equation}
We use the same approach as for the randomly-weighted ER graphs; that is, we add self-connections 
to the nodes and random weights to the corresponding adjacency matrix. Under these conditions, the 
average degree is written as
\begin{equation}
    \langle k \rangle =1+(N-1)F(r).
    \label{eqkrg}
\end{equation}

\begin{figure}
\centering
\includegraphics[scale=0.4]{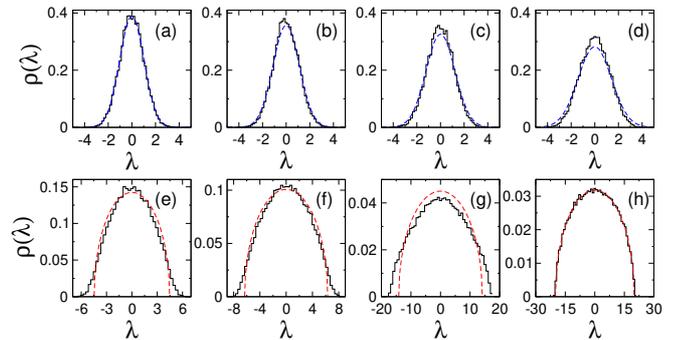}
\caption{
Spectral density of randomly-weighted random geometric graphs of size $N=200$ 
and (a) $\langle k \rangle =1.05$, (b) $\langle k \rangle =1.25$, (c) $\langle k \rangle =1.5$,
(d) $\langle k \rangle =2$, (e) $\langle k \rangle =10$, (f) $\langle k \rangle =20$, 
(g) $\langle k \rangle =100$, and (h) $\langle k \rangle =200$. 
Blue and red dashed lines correspond to Eqs.~(\ref{den1}) and~(\ref{den2}), respectively. 
The eigenvalues of 100 adjacency matrices were used to construct each histogram.}
\label{Fig09RG}
\end{figure}
\begin{figure}
\centering
\includegraphics[scale=0.4]{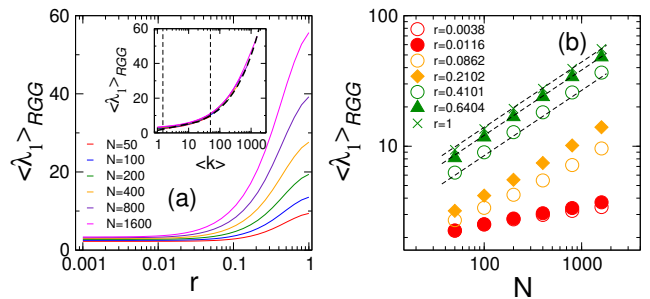}
\caption{
(a) The average largest eigenvalue $\bra \lambda_{1}\ket$, of randomly-weighted random geometric graphs,
as a function of the connection radius $r$. 
Inset: $\bra \lambda_{1}\ket$ as a function of the average degree $\langle k \rangle$.
(b) $\bra \lambda_{1}\ket$ as a function of the graph size 
$N$ for different values of $r$, as indicated in the panel. 
Dashed lines in the inset of panel (a) and in panel (b) correspond to Eq.~(\ref{lmb1}).
Each symbol was calculated by averaging over $10^6/N$ random graphs.}
\label{Fig10RG}
\end{figure}
\begin{figure}
\centering
\includegraphics[scale=0.4]{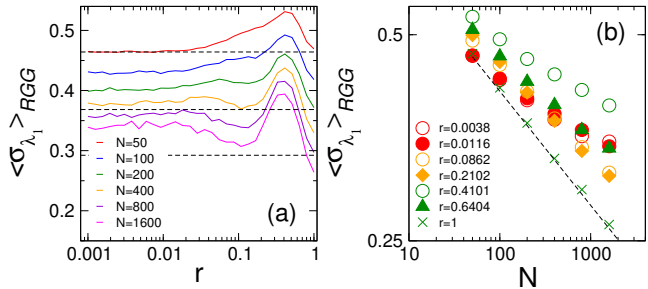}
\caption{
(a) The average variance of the largest eigenvalue $\bra \sigma_{1}\ket$, of randomly-weighted 
random geometric graphs, as a function of the connection radius $r$. (b) $\bra \sigma_{1}\ket$ 
as a function of the graph size $N$ for different values of $r$, as indicated in the panel. 
Dashed lines in (a) correspond to $N^{-1/6}$ for $N=100,400$ and $1600$. 
Dashed line in (b) is $N^{-1/6}$
Each symbol was calculated by averaging over $10^6/n$ random graphs.}
\label{Fig11RG}
\end{figure}

\subsection{Results}

\begin{figure}
\centering
\includegraphics[scale=0.41]{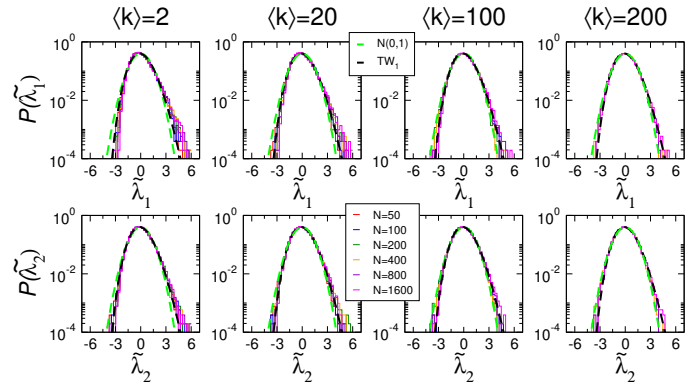}
\caption{
Top panels: Distributions of the largest eigenvalue $P(\lambda_1)$ of randomly-weighted 
random geometric graphs for different values of $\langle k \rangle$, as indicated at the 
top of the panels.
Bottom panels: Distributions of the second largest eigenvalue $P(\lambda_2)$.
Each panel includes histograms for several graph sizes $n \in [50,400]$. 
Black and green dashed lines correspond to the Tracy-Widom distribution of 
type 1 and normal distributions, respectively. 
Each histogram was constructed from $10^6/n$ random graphs.}
\label{Fig12RG}
\end{figure}

\begin{figure}
\centering
\includegraphics[scale=0.3]{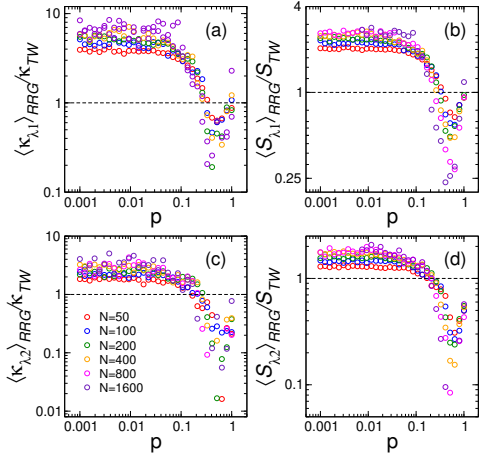}
\caption{
(a) [(c)] Excess kurtosis and (b) [(d)] skewness, normalized to the corresponding values of the 
Tracy-Widom distribution of type 1, for the largest eigenvalue $\lambda_1$ [for the 
second largest eigenvalue $\lambda_2$] as a function of the connection radius $r$ of 
randomly-weighted random geometric graphs of sizes $n \in [50,400]$. 
Each symbol was computed by averaging over $10^6/n$ random graphs.}
\label{Fig13RG}
\end{figure}

\begin{figure}
\centering
\includegraphics[scale=0.3]{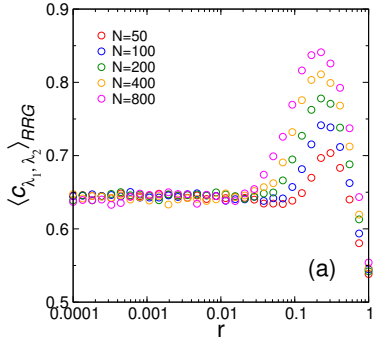}
\caption{
Correlation coefficient between the largest and the second largest eigenvalues 
as a function of the connection radius $r$ of randomly-weighted 
random geometric graphs of sizes $n \in [50,400]$.
Each symbol was computed by averaging over $10^6/n$ random graphs.}
\label{Fig14RG}
\end{figure}

Our results for randomly-weighted RGGs are presented in Figs.~\ref{Fig09RG}--\ref{Fig16RG}.
Note that Figs.~\ref{Fig09RG}--\ref{Fig16RG} for randomly-weighted RGGs are equivalent to 
Figs.~\ref{Fig01ER}--\ref{Fig08ER} for randomly-weighted ER graphs, respectively. 
Indeed, from Figs.~\ref{Fig09RG}--\ref{Fig16RG} we draw similar conclusions as those already 
discussed in the main text for randomly-weighted ER graphs:
\begin{itemize}

\item[{\bf (i)}] 
The spectral density of randomly-weighted RGGs has a Gaussian shape for
$\langle k \rangle < 2$ and a semicircular shape for $\langle k \rangle > 2$ as given by
Eqs.~(\ref{den1}) and~(\ref{den2}), respectively; see Fig.~\ref{Fig09RG}.
However, significant deviations are observed for given values of $\langle k \rangle$; see 
e.g.~Fig.~\ref{Fig09RG}(g), which may need further exploration.

\item[{\bf (ii)}]  
The average largest eigenvalue $\bra \lambda_{1}\ket$ of randomly-weighted 
RGGs depends on the average degree as given by Eq.~(\ref{lmb1}); see the inset of 
Fig.~\ref{Fig10RG}(a) and Fig.~\ref{Fig10RG}(b).

\item[{\bf (iii)}] The average variance of the largest eigenvalue exhibits a power-law dependence on network size; see Fig.~\ref{Fig11RG} (b), with an exponent that varies with the connection radius. This variation becomes particularly noticeable for $r>0.1$ as reflected in the bell-shaped behavior shown in Fig.~\ref{Fig11RG} (a).

\item[{\bf (iv)}] 
The distributions of the largest eigenvalue $P(\lambda_1)$ and of the second 
largest eigenvalue $P(\lambda_2)$ of randomly-weighted RGGs approach a Tracy-Widom 
distribution of type 1 for $r\to \sqrt{2}$ (or $\langle k \rangle \gg 1$); see Figs.~\ref{Fig12RG} 
and~\ref{Fig13RG}.

\item[{\bf (vi)}] 
The correlation coefficient between the largest and the second largest eigenvalues 
of randomly-weighted RGGs displays the PE--to--GOE transition as a function of $r$.
That is, $c_{\lambda_1,\lambda_2}\approx 0.65$ for small $r$, while 
$c_{\lambda_1,\lambda_2}\approx 0.53$ for $r\to \sqrt{2}$; see Fig.~\ref{Fig14RG}.

\item[{\bf (vii)}] 
The distribution of the normalized distance between the first and second eigenvalues $P(s)$ 
and the distribution of the ratio between higher consecutive eigenvalues spacings $P(r)$ of 
randomly-weighted RGGs displays the PE--to--GOE transition as a function of $\langle k \rangle$.
Moreover, for small [large] values of $\langle k \rangle$, the shapes of $P(s)$ and $P(r)$ are 
well described by $P_{\tbox{PE}}(s)$ and $P_{\tbox{PE}}(r)$ [$P_{\tbox{GOE}}(s)$ and 
$P_{\tbox{GOE}}(r)$], respectively; see Fig.~\ref{Fig15RG}. 

\item[{\bf (viii)}] 
The IPR distributions of all eigenvalues display a clear delocalization transition as $\langle k \rangle$ 
increases; see Fig.~\ref{Fig16RG}.
The IPR distributions of extreme eigenvalues (largest, second largest, and smallest) exhibit 
similar localization characteristics, in contrast with those for the bulk eigenvectors.
Specifically, for small values of $\langle k \rangle$, the IPR distributions show a pronounced peak at 
$\mbox{IPR}=3$, a signature of strongly localization and a characteristic of the PE. While for large 
values of $\langle k \rangle$ the IPR distributions follow a Gaussian-like bell shape centered at 
$\mbox{IPR}\propto N$, indicating delocalization which is the main characteristic of the GOE.

\end{itemize}

\begin{figure}
\centering
\includegraphics[scale=0.4]{Fig15RG.eps}
\caption{
Top panels: Distribution of the normalized distance between the largest 
and second-largest eigenvalues $P(s)$ of randomly-weighted random geometric graphs 
for different values of $\langle k \rangle$, as indicated at the top of each panel.
Bottom panels: Distribution of the ratio between higher consecutive eigenvalues spacings $P(r)$. 
Each panel displays histograms for different graph sizes, $n \in [50,400]$. 
In the top panels, black and green dashed lines correspond to Eqs.~(\ref{psp}) and~(\ref{psg}), respectively. 
In the bottom panels, black and green dashed lines correspond to Eqs.~(\ref{prp}) and~(\ref{prg}), respectively. 
Each histogram was constructed from $10^6/n$ random graphs.}
\label{Fig15RG}
\end{figure}
\begin{figure}
\centering
\includegraphics[scale=0.4]{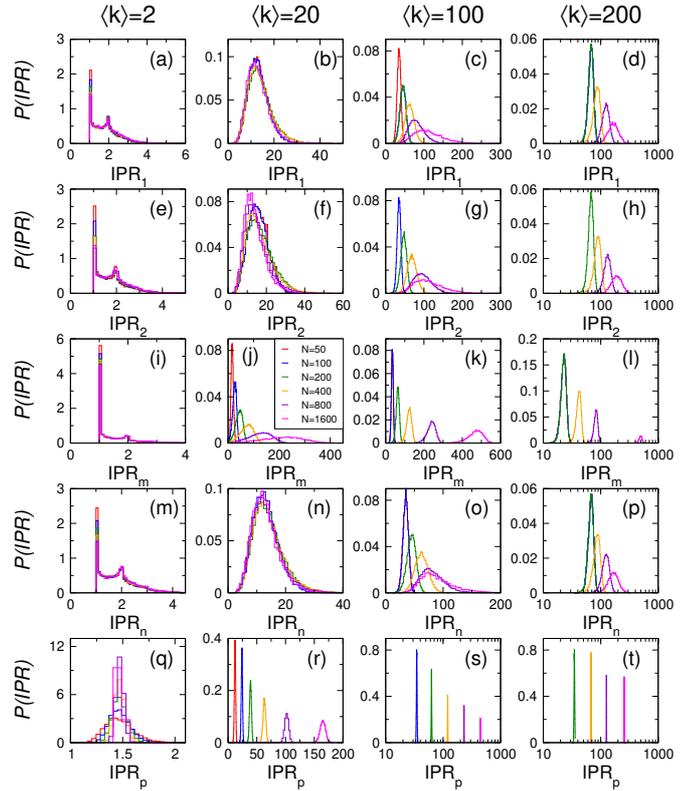}
\caption{
Distributions of inverse participation ratios $P(\mbox{IPR})$ of eigenstates of randomly-weighted 
random geometric graphs of sizes $N \in [50,800]$. 
Each column corresponds to a fixed value of $\langle k \rangle$.
Rows are $P(\mbox{IPR})$ of eigenstates corresponding to: 
(a-d) the largest eigenvalue, (e-h) the second largest eigenvalue, 
(i-l) the central eigenvalue $(m=N/2)$, (m-p) the smallest eigenvalue, and (q-t) the average IPR over the full spectrum.
}
\label{Fig16RG}
\end{figure}

\begin{acknowledgments}
C.T.M.-M. Thanks for the support from CONAHCYT (CVU No.~784756).
J.A.M.-B. Thanks for the support from VIEP-BUAP (Grant No.~100405811-VIEP2025), Mexico.
\end{acknowledgments}

\end{document}